**Title**: Plant sesquiterpene lactones

**Authors**: Olivia Agatha[1], Daniela Mutwil-Anderwald[1], Jhing Yein Tan[1], Marek Mutwil[1]*
   1. School of Biological Sciences, Nanyang Technological University, 60 Nanyang Drive, Singapore, 637551, Singapore

**Abstract**
Sesquiterpene lactones (STLs) are a prominent group of plant secondary metabolites predominantly found in the Asteraceae family and have multiple ecological roles and medicinal applications. This review describes the ecological significance of STLs, highlighting their roles in plant defense mechanisms against herbivory and as phytotoxins, alongside their function as environmental signaling molecules. We also cover the substantial role of STLs in medicine and their mode of action in health and disease. We discuss the biosynthetic pathways and the various modifications that make STLs one of the most diverse groups of metabolites. Finally, we discuss methods in identifying and predicting STL biosynthesis pathways.

**Introduction**
Sesquiterpene lactones (STLs) are a large group of plant secondary metabolites with a high diversity in structure and biological activity. They are a terpenoid subgroup characterized by a 15-carbon backbone structure containing at least one lactone group with a carbonyl moiety. The 15C backbone can be rearranged in several ways, is mostly cyclic, and is further decorated with different functional groups, which results in a large variety of sesquiterpenoids. The main subtypes are germacranolides, guaianolides, and eudesmanolides [1]. About 5,000 known STL structures are found in the Asteraceae family, the second-largest plant family after Orchidaceae. Asteraceae contain about 23,000 plant species and is considered the most advanced family of dicots in the plant kingdom. The wide chemical diversification of STLs is interlinked with plant speciation in Asteraceae, and thus, STLs have been used as chemotaxonomic markers for classification since the 1950s [2,3]. STLs are also widely distributed in other plant families, including Acanthacea, Euphorbiaceae, Lamiaceae, Lauraceae, Magnoliaceae, Orchidaceae, and even in liverworts [1].

*Ecological function in herbivory*
In plants, the highest amounts of STLs have been found in specialized cells, such as capitate glandular trichomes, resin ducts, or laticifers [4,5], which corroborated their suggested role as anti-herbivory agents [6]. Leaves of *Tithonia diversifolia* (Asteraceae), for example, contain varying levels of STLs over the year, and feeding of *Chlosyne lacinia* larvae stops when levels are high (0.3% vs. 0.1% relative to leaf dry weight) [7]. Some STLs can also affect insect development by reducing the growth rate, e.g., glaucolide A from

*Vernonia*, cnicin from *Centaurea maculosa,* and 8ß-sarracinoyloxy-cumambranolide from *Helianthos* [6,8,9]. STLs were shown to directly affect insect mortality, most likely by reducing *in vivo* glutathione levels, which leads to oxidative damage of cellular membranes in insect larvae [10]. The metabolism of the herbivore can, however, in some cases, modify the toxicity of the STL, as in the case of taraxinic acid β-D-glucopyranosyl ester from dandelion [11]. With its manifold roles in plant defense, STLs most likely contributed to the evolutionary success of the Asteraceae family.

*Ecological function as phytotoxin*
Apart from their anti-herbivory roles, several STLs were long known to act as potent phytotoxins, i.e., growth inhibitors of competing plants, thus creating an ecological advantage [12]. One of the first identified STLs was parthenin isolated from the tropical aggressive weed *Parthenium hysterophorus* (Asteraceae), which affects the growth of crops, other Asteraceae, and even its species [13]. Sunflower leaves also possess a number of STLs, e.g., germacranolides and guaianolides, that inhibit germination or root growth of other species [14]. Even artemisinin and its synthetic analogues is a potent root growth inhibitor of *Lolium multiflorum* (Poaceae) at a low micromolar concentration of 3.5 µM [15]. However, the relevance of growth inhibition in the field remains a matter of debate, as STLs are not easily released into the environment (mainly by leaching due to rain or decomposition). Thus, phytotoxin concentration may be too low to affect neighboring plants [1,16]. Nevertheless, STLs are promising alternatives to synthetic herbicides. Recently, a new strigolactone analogue based on STLs was developed based on eudesmanolides with potential use as preventive agrochemicals for broomrape control [17].

*Ecological function as an environmental signaling molecule*
Although the highest amounts of STLs have been found in specialized cells, lower amounts were also found in inner plant tissues, including the root, e.g., in sunflower *(Helianthus annuus)*, false ragweed (*Parthenium hysterophorus*), and chicory (*Cichorium intybus* L.) [18–20]. Interestingly, some of these endogenous STLs, such as costunolide and dehydrocostus lactone, promote germination and growth of parasitic plant species such as *Orobranche* and *Striga* at low micromolar concentrations when exuded from sunflower roots [21], but irreversibly inhibit germination at higher concentrations above 100 mM [22]. This dual action is similar to strigolactones, a group of phytohormones structurally related to STLs.

*STLs as signaling molecules*
Some endogenous STLs interact with hormones, which has profound implications on the physiology of the plant. For example, when exposed to light, niveusin C and 15-hydroxy-3-dehydro-desoxyfructicin inhibit auxins at micromolar concentrations in leaves of sunflower seedlings by binding to a thiol group in the auxin receptor [23]. Inhibition of auxin leads to slower elongation on the illuminated side of the hypocotyl, and thus bending towards the light. In line with this role in phototropism, it was shown in sunflower that the endogenous levels of 8-epixanthatin were three-fold increased in the illuminated side than in the shade [24]. The inhibiting effects of 8-epixanthatin and other STLs (such as costunolide and dehydrocostus lactone) have been recently investigated on a molecular level in sunflower from the onset of germination and in response to light [18]. The ability to inhibit the polar auxin transport was shown by Toda et al. (2019)[25], who studied the growth inhibition of dehydrocostuslactone on etiolated pea seedlings. The effects correlated with the reduction of gene expression for plasma membrane-located proteins needed for auxin transport.

*Usage in medicine*

Apart from its roles in plant biology, STLs display a broad spectrum of biological activities relevant to human health, leading to intense research efforts in the past 20 years. Some compounds have been identified to be highly specific and thus suitable as therapeutics. The most well-known is artemisinin, an STL isolated from *Artemisia annua*, which has been used since the early 1980s for treating malaria caused by the parasite *Plasmodium falciparum* [26]. Remarkably, no severe side effects have been reported for this drug or its more water-soluble derivatives, such as artemeter, artesunate, or artemisone [27,28]. Artemisinin is a cadinanolide with a 1,2,4-trioxane ring system, and its mode of action is based on the selective uptake of the compounds by infected blood cells, consequently causing accumulation of damaged, ubiquitinated proteins in the parasite [29].

Several STLs offer promising lead structures to design novel drugs against cancer [30,31] and inflammation [32,33]. Apart from its antimalarial activity, artemisinin was found to possess anti-cancer activity, a topic of intense current research [34,35]. Other well-studied STLs that display anti-cancer activities are parthenolide (a germacranolide), and thapsigargin (a guaianolide). Parthenolides are present in feverfew (*Tanacetum parthenium*) and display inhibitory effects on the transcription factors belonging to NF-κB and STAT families, both involved in apoptosis, cell proliferation, and inflammation. NF-κB regulates over 150 inflammatory genes, including many involved in cell proliferation, and mediates immune response by controlling the expression of proinflammatory mediators, including cytokines [36]. Importantly, parthenolide (and its more water-soluble derivatives) has no toxic effect on normal cells and has been used in several preclinical studies [37,38] and also in phase 1 trials for patients with cancer [39–41]. Moreover, parthenolide is one of the few small molecules that specifically target cancer stem cells, considered the "initiating" cells of the tumor [36,42].

Thapsigargin was first isolated in 1978 from the plant *Thapsia garganica* to characterize its skin irritant properties [43]. Further studies revealed that thapsigargin can induce apoptosis by inhibiting the Sarco/Endoplasmic Reticulum Calcium ATPase (SERCA) [44]. Due to its cytotoxicity to normal cells, thapsigargin was initially not considered for development as an anti-cancer drug. However, by adding a polypeptide designed to be cleaved by a protease specifically produced in the extracellular space of metastatic prostate cancer cells, the prodrug mipsagargin (G-202) was synthesized [45]. After cleavage, the released hydrophobic thapsigargin enters the cancer cells and promotes Ca2+-mediated cell death. Phase II clinical trials using mipsagargin against hepatocellular carcinoma showed that it is relatively well tolerated and promotes prolonged patient disease stabilization [46]. For a comprehensive review of all STLs, including artemisinin and derivatives, micheliolide (ACT001), thapsigargin/mipsagargin (G-202), parthenolide, arglabin and atractylenolide 1 that have undergone or are undergoing cancer clinical trials, please refer to Cheikh et al. 2022 [47], and for the bioactivity of parthenolide, please refer to the excellent review by Freund et al. 2020 [48].

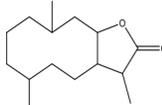

Figure 1. Major types of sesquiterpene lactones (STLs) in plants. The six major types of STLs are represented by the structural backbone and specific examples, together with the plant species where the molecule is found. References are: [1] Krishna et al., 2018 [49], [2-4] Pareek et al. 2011 [50], [5, 7, 9, 10, 20] Cheikh et al. 2022 [47], [6] Karthikeyan et al. 2016 , [8] Tobyn et al. 2011 [51], [11] Photo by Mauricio Mercadante, available from eol.org/pages/5118025, [12] Branquinho et al. 2020 [52], [13] Photo by Alex Popovkin, available from https://eol.org/pages/47145431. [14] Schefer et al. 2017 [53], [15-16] Schmidt 2023 [54], [17] Bezuneh 2015 [55], [18] Kaur et al. 2021 [56], [19] Nam et al. 2015 [57].

*Biological mode of action*

The biological activities of most STLs are primarily attributed to the lactone structures with a double bond next to the carbonyl group, such as the $\alpha,\beta$-unsaturated $\gamma$-lactone moiety and the α-methylene-γ-lactone group (αMγL). The αMγL group, an oxygen-containing ring structure with a carbonyl function, causes the alkylation of thiol groups commonly found in proteins, thereby inhibiting their function. The alkylation power of αMγL also acts on transcription factors and enzymes, causing steric and chemical changes to the target by Michael reactions [58–60]. There are exceptions from Michael-type additions, such as for artemisinin and thapsigargin, where an exocyclic methylene is responsible for the bioactivity. Interestingly, analogs of 10-Deoxoartemisinin substituted at C-3 and C-9 were up to 60 times more active [61]. This implies that small modifications by chemical design can dramatically increase the therapeutic potential of STLs.

**Types of sesquiterpene lactones**

STLs are derived from the mevalonic acid pathway, and their biosynthesis involves several enzymes and intermediate compounds. The most widespread structural groups are germacranolides, heliangolides, guaianolides, pseudoguaianolides, and eudesmanolides (Figure 1). Minor structural groups include elemanolides, eremophilanolides, and xanthanolides. These groups can be categorized based on their carbocyclic backbone. Germacranolides and heliangolides have a 10-membered carbon ring, xanthanolides have a 7-membered ring, elemanolides are 6-membered ring compounds, guaianolides, and pseudoguaianolides are 5/7-bicyclic compounds, while eudesmanolides and eremophilanolides belong to a group of 6/6-bicyclic compounds.

Germacranolides are the largest group of naturally occurring STLs and the most significant with regard to human health. Their ten-membered molecular core is fused with a five-membered γ-lactone. A prototypical example of a germacranolide is costunolide. Like parthenolide, costunolide can induce apoptosis via reactive oxygen species (ROS) generation in various types of cancer cells [62], and also shows anti-inflammatory and antioxidant properties [63].

Reports on the pharmacological effects of heliangolides are still scarce. The furanoheliangolide goyazensolide has shown strong inhibiting activities (IC50 of 0.6 µM) of c-Myb-dependent gene expression in cancer cells [64]. The transcription factor c-Myb is a proto-oncogene that regulates the expression of genes involved in proliferation, cell survival, and differentiation and is an interesting therapeutic target. Other representative examples with pharmacological activities include lychnopholide, which can be extracted from *Lychnophora trichocarpha* [65], and diacethylpiptocarphol from *Vernonia scorpioides* [53,66].

Guaianolides have 7- and 5-membered rings and a methyl group at the C-4 position. Two subclasses of guaianolides are the 6,12-guaianolides and 8,12-

guaianolides. In particular, α-methylene guaianolides and guaianolides containing an α,β-unsaturated carbonyl moiety show biological activities [67]. An example is dehydrocostus lactone, an effective anti-inflammatory agent inhibiting the expression of key inflammatory markers in activated macrophages [68]. The structural analogue with an α-methylene-γ-butyrolactone moiety that lacks a double bond displayed minimal anti-inflammatory activity, indicating the significance of the α,β-unsaturated carbonyl group in dehydrocostus lactone and underscores the importance of specific structural features responsible for pharmacological effects [69]. A novel anti-inflammatory and anti-cancer guaianolide is micheliolide, which was synthesized based on parthenolide and shows high stability and effectiveness, particularly as a Micheal adduct [70,71]. Micheliolide (ACT001) is currently administered with immunotherapy and radiotherapy treatments in phase 2 trials [47].

Pseudoguaianolides have 7- and 5-membered rings and a methyl group at the C-5 position. Anticancer activities have been evidenced *in vitro* and *in vivo* with different tumor models for britannin, found in different Inula species, and helenalin from *Arnica montana* [72,73]. Helenalin acetate suppresses Myb-dependent gene expression similar to goyazensolide [64].

Eudesmanolides have a bicyclic 6-6-fused ring system fused to the butyrolactone group. Several eudesmanolides from the flowers of *Sphagneticola trilobata* (L.) Pruski show significant anti-proliferative and anti-inflammatory activities [74]. Alantolactone and isoalantolactone isolated from *Inula helenium* are other examples of eudesmanolides with anti-inflammatory properties [75].

Artemisinin is a unique STL from Artemisia annua. It does not belong to any of the general subgroups mentioned above as it has a sesquiterpene trioxane lactone with an endoperoxide bridge. The endoperoxide bridge is crucial for its antimalarial and neuron growth-stimulating properties [76].

**The biosynthesis of major classes of STLs**

All STLs are derived from farnesyl diphosphate (FPP), which is derived from dimethylallyl diphosphate (DMAPP) and isopentenyl diphosphate (IPP) (Figure 2). DMAPP and IPP can be either biosynthesized from the mevalonate (MVA) or the 2-C-methylerythritol 4-phosphate (MEP) pathway. The MVA pathway occurs in the cytoplasm [77], while the MEP pathway occurs in the plastid [78]. Both pathways generate the five-carbon monomer IPP and its isomer DMAPP [79], which are the precursors (isoprene units) for the biosynthesis of terpenoids.

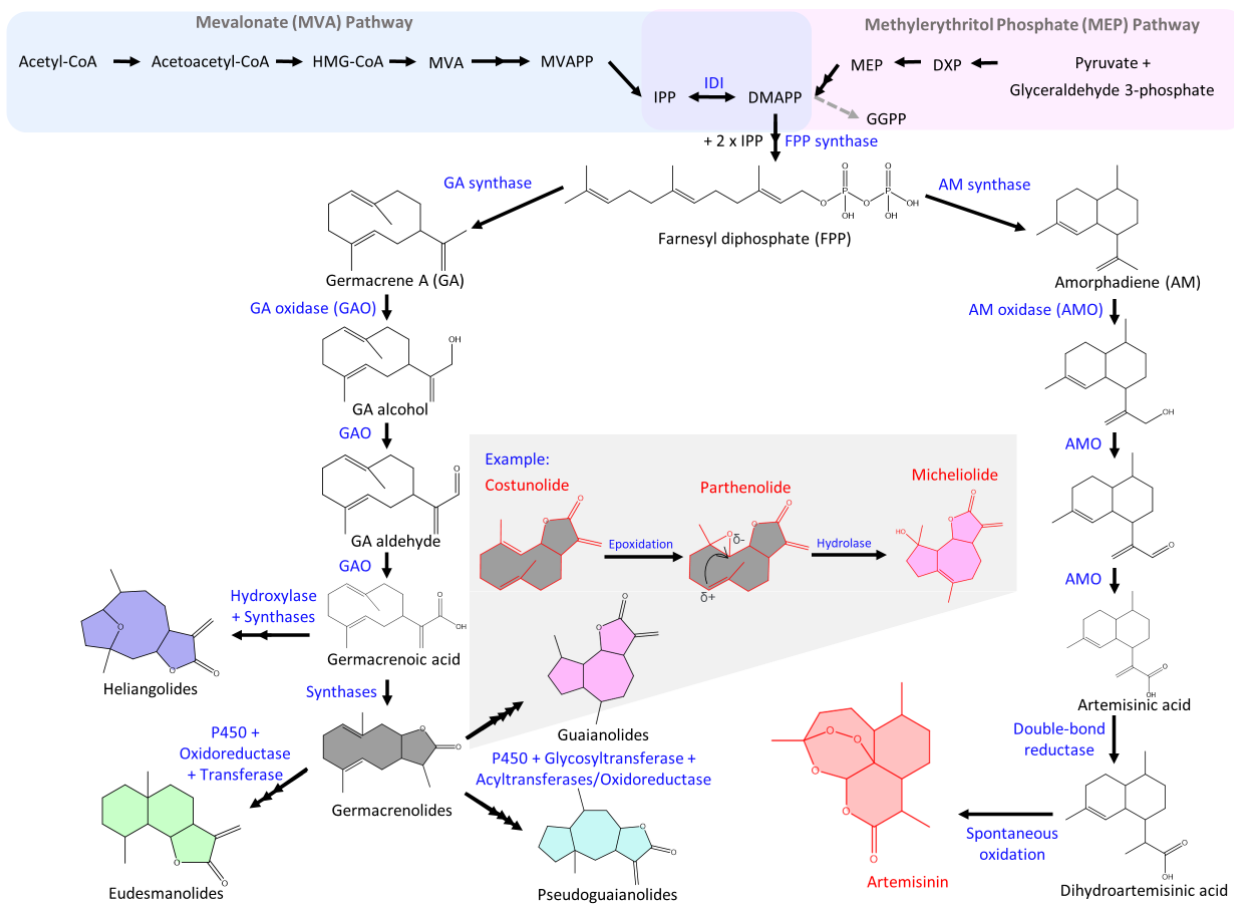

**Figure 2. Biosynthetic pathways of sesquiterpene lactones.** The abbreviations are: Acetyl coenzyme A (Acetyl-CoA); 3-hydroxy-3-methylglutaryl-CoA (HMG-CoA); Mevalonate (MVA), Mevalonate diphosphate (MVAPP); Isopentenyl diphosphate (IPP); Isopentenyl diphosphate isomerase (IDI); Dimethylallyl diphosphate (DMAPP); 2-C-methyl-D-erythritol 4-phosphate (MEP); Geranylgeranyl diphosphate (GGPP); 1-deoxy-d-xylulose 5-phosphate (DXP); Farnesyl diphosphate (FPP); Germacrene A (GA); Germacrene A oxidase (GAO); Amorphadiene (AM); Amorphadiene oxidase (AMO); Cytochrome P450 (P450).

The MVA pathway begins with the Claisen condensation of acetyl-CoA to acetoacetyl-CoA by acetoacetyl-CoA thiolase (AACT) [80]. Next, the acetoacetyl-CoA undergoes an aldol condensation with another acetyl-CoA to form 3-hydroxy-3-methylglutaryl-CoA (HMG-CoA) with the help of 3-hydroxy-3-methylglutaryl-coenzyme A synthase (HMGS), $Fe^{2+}$ and plastid quinone. Then, a rate-limiting step of conversion of HMG-CoA to mevalonate by HMG-CoA reductase (HMGR) takes place. Afterward, HMG-CoA undergoes two ATP-dependent phosphorylation steps catalyzed by mevalonate kinase (MK) and phosphomevalonate kinase (PMK), and one ATP-driven decarboxylation catalyzed by mevalonate diphosphate decarboxylase (MVD), which results in IPP. IPP isomerization to DMAPP is catalyzed by isopentenyl diphosphate isomerase (IDI).

Although a minor contributor to FPP, the MEP pathway can also produce FPP. The MEP pathway occurs through the condensation of a pyruvate and glyceraldehyde 3-phosphate (GAP) to form 1-deoxy-d-xylulose 5-phosphate (DXP) with the help of DXP synthase. The DXP is then reduced by DXP reductoisomerase to form the intermediate 2-C-methyl-D-erythritol 4-phosphate (MEP). Through seven types of enzymes, MEP finally produces IPP and DMAPP, which in the MEP pathway are mainly converted to geranylgeranyl pyrophosphate (GGPP) and subsequently produce gibberellins and diterpenes [81].

FPP is synthesized by adding two units of IPP to DMAPP via condensation catalyzed by farnesyl diphosphate synthase (FPPS) [82]. FPP is a necessary precursor for several classes of crucial metabolites other than STLs, such as homoterpenes, triterpenes, sterols, brassinosteroids, and polyprenols [83]. For sesquiterpene lactones, FPP has the potential to undergo various cyclization pathways and oxidative transformations, leading to the formation of either cis (Z) or trans (E)-fused lactones [84]. The cyclization of FPP is usually mediated by germacrene A synthase (GAS) to form germacrene A, a common intermediate for most of the sesquiterpene lactone subclasses. Germacrene A undergoes carbocyclic rearrangement and oxidation before lactonization with a lactone ring to produce germacranolides [84]. Another example of a cyclization enzyme of FPP is amorphadiene synthase, which forms amorphadiene, a precursor to artemisinin [85,86].

**Diversification of STLs through additional decorations**
Different sesquiterpene synthases and downstream modifications such as oxidation, halogenation, acetylation, glycosylation, and glucosylation on intermediate compounds give rise to a diverse array of STL structures (Figure 3). The most common modification by tailoring enzymes is oxidation, but STLs can also be decorated by sugars, halogens, and other groups.

*Oxidation*
Oxidation of sesquiterpenes, including STL, is most commonly catalyzed by cytochrome P450 oxidases (other enzymes are peroxidases, dehydrogenases, and oxidoreductases). In Asteraceae, all known cytochrome P450 enzymes involved in sesquiterpene metabolism are members of the CYP71 family. Some examples include costunolide synthase, eupatolide synthase, artemisinic alcohol dehydrogenase, and germacrene A oxidase. In some cases, one enzyme is responsible for more than one oxidation step. For example, germacrene A oxidase adds an -OH group to germacrene A, followed by oxidizing this hydroxyl group to an aldehyde and finally to carboxylic acid, hence forming germacrene A acid [87]. Amorphadiene monooxygenase behaves similarly to germacrene A oxidase and converts amorphadiene into artemisinic aldehyde through two oxidation steps [88,89]. Aldehyde dehydrogenase 1 (ALDH1) subsequently oxidizes the hydroxyl group to a carboxylic acid functional group to form artemisinic acid [88]. Another example of an oxidation step is costunolide synthase, which adds an -OH group to germacrene A acid. After the addition of a -COOH group by the germacrene A oxidase, the compound spontaneously forms a lactone ring, which results in costunolide [90].

Oxidation steps can also occur after the lactone formation. In many cases, oxidation of a compound allows subsequent modifications by other types of enzymes. For example, Tp8879 kauniolide synthase from *Tanacetum parthenium* adds an -OH group to costunolide to form 3α-hydroxycostunolide, followed by a ring closure reaction, which results in kauniolide formation [91]. Another example is parthenolide synthase from *Tanacetum parthenium,* which adds an epoxy group to costunolide, resulting in parthenolide, which can then be further oxidized by CYP71 oxidases to form 9α-hydroxyparthenolide (containing -OH), 8α,9α-epoxyparthenolide (containing -O-), and parthenolide-9-one (containing =O) [92].

*Halogenation*
Most halogen-containing natural products come from marine environments, although some can also be found on land. Halogenation can be catalyzed by five different classes of enzymes: heme iron-dependent haloperoxidases, vanadium-dependent haloperoxidases, non-heme iron-dependent halogenases, flavin-dependent halogenases, and nucleophilic halogenases [93]. Marine halogen-containing sesquiterpenes include prepacifenol, prepacifenol epoxide, and pacifenol found in the red alga Laurencia [94]. There are also halogenated STLs, including Linderagalactone from *Lindera sp.* [95], Purpuride B from *Talaromyces minioluteus* (*Penicillium minioluteum*) [96], and laurenolide A and laurenolide B from *Palisada intermedia* [97]. Land plants also produce halogenated STLs. For example, plants of the genus Centaurea produce chlorinated guaianolides, including chlorohyssopifolin A (Figure 3), B, C, D, E, and linichlorin A and C [98,99]. The halogenation of these guaianolides is proposed to be performed by flavin-dependent halogenases, specifically FADH2-dependent halogenases, which act on an -OH group previously introduced by a hydroxylase (Figure 3). *Cynara cornigera* has also been found to contain two chlorinated guaianolides, named cornigeraline A and 13-chlorosolstitialine [100]. Other chlorinated heliangolides include lacinolide B extracted from *Viguiera laciniata* [101], and a chlorinated form of mikanolide (a germacranolide with two lactone rings) isolated from *Mikania cordata* [102]. Unfortunately, the enzymes responsible for the halogenation of these compounds have not yet been found.

*Acetylation*
Acetyltransferases catalyze the transfer of acetyl groups ($CH_3CO-$). For example, trichothecene 3-O-acetyltransferase (TRI101) catalyzes the acetylation of a free C3 hydroxyl group of isotrichodermol and other trichothecene mycotoxins in *Fusarium*

*sporotrichioides* and in this way protects the fungus from the toxic end products of the sesquiterpene biosynthesis pathway [103]. Before the mycotoxins are secreted by the fungus, the acetyl group is removed.

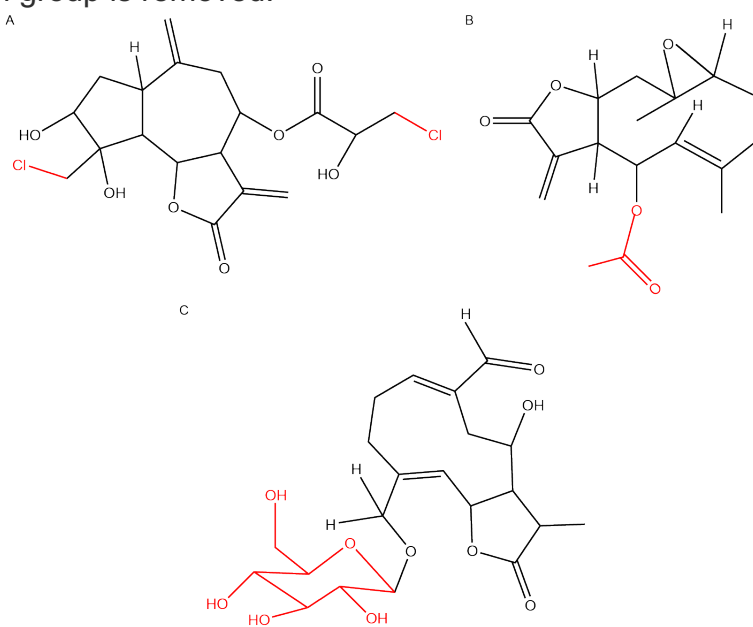

**Figure 3. Examples of Modifications of Sesquiterpene Lactones.** (A) The chlorinated guaianolide Chlorohyssopifolin A; (B) The acetoxylated germacranolide Pyrethrosin. (C) The glucosylated germacranolide 15-O-β-D-glucopyranosyl-11β,13-dihydrourospermal A;

In some Asteraceae, acetyltransferases are responsible for the acetylation of guaianolides by transferring an acetyl moiety from acetyl CoA to one of the -OH groups of the guaianolide, resulting in an acetylated guaianolide, e.g. in *Centaurothamnus maximus* [99]. Furthermore, there are acetoxylated germacranolides, e.g., β-cyclopyrethrosin, pyrethrosin (Figure 3), dehydro-β-cyclopyrethrosin, and chrysanolide that can be found in *Tanacetum cinerariifolium* [104], of which pyrethrosin shows clear phytotoxic, antibacterial, and antifungal properties. In addition, there are acetoxylated heliangolides found in *Viguiera laciniata*[101], and acetoxylated germacranolides and eudesmanolides also exist in *Dicoma sp* [105]. However, the exact enzymes responsible for the acetoxylation have not been discovered.

*Glycosylation and Glucosylation*
There are several examples of sesquiterpene lactone glucosides in plants, e.g., isolated from *Ixeris dentata* [98] and *Sonchus asper, e.g. 15-O-β-D-glucopyranosyl-11β,13-dihydrourospermal A* (Figure 3)*.* In addition, a particular form of glucosylated STL exists in *I. salsoloides*, named  eupatolide-13-*O*-β-D-glucopyranoside [107]. A glycosylated germacranolide,  (11-R)-6-*O*-β-D-glucosyl-11–13 dehydro-tatridin-B, was found in *Tanacetum cinerariifolium* [104]. Moreover, four sesquiterpene lactone glycosides have been extracted from the roots of *Crepis capillaris* [108]. Seeds of *Carpesium macrocephalum* have also been found to contain sesquiterpene lactone glycosides [109]. A eudesmanolide glycoside called 11,13-dihydro-3-O-(β-digitoxopyranose)-7αhydroxy eudasman-6,12-olide was extracted from *Sphaeranthus indicus* Linn [110].

**Prediction of SLT Biosynthesis Pathways**
As STL can have important therapeutic values, a more economically feasible production and development of new varieties would be of high value. However, many STLs are difficult to synthesize chemically [111]. It is also time-consuming and expensive to extract STLs from plants [112]. Consequently, there is a significant effort to elucidate the biosynthetic pathway of STLs to produce these compounds in more suitable hosts, such as bacteria and fungi [113]. A most famous example is the near-total biosynthesis of artemisinin in yeast [114]. Unfortunately, the exact biosynthetic pathway of most STLs is still unknown, but we have several tools to predict these pathways. These methods comprise sequence- and expression-based approaches.

*Sequence-based Pathway Prediction*
Sequence similarity approaches are often the first step to predict new STL synthases. For example, BLASTx correctly identified two new sesquiterpene synthases from *Polygonum minus* as the amino acid sequences of these two candidates were strongly similar to drimenol synthase from *Persicaria hydropiper* [115]. Similarly, two new P450s were identified in *C. intybus* due to their sequence similarity to germacrene A oxidase (CiGAO) and costunolide synthase (CiCOS) [116]. Tools such as PlantiSMASH use hidden Markov models (HMMs) or Basic Local Alignment Search Tool (BLAST) to indicate genes associated with the STL biosynthesis [117]. If a genome is absent, one can conduct short-read RNA sequencing of the different organs of the organism and assemble a transcriptome [118]. Ensemble Enzyme Prediction Pipeline (E2P2), which utilizes homology to annotate functions, can then be used to predict enzymes. To exemplify how E2P2 can be used, we searched for STL transcripts in 100 representative species of Archaeplastida and identified EC (Enzyme Commission) numbers starting with 4.2.3, which represents Carbon-oxygen lyases acting on phosphates. Within this group, we observed 88 sesquiterpene cyclase classes. We also found 2801 sesquiterpene synthase candidate transcripts in 94 species (Figure 4). From these synthase candidates, several are present exclusively in certain plant clades. Trichodiene, (-)-gamma-cadinene, pentalenene, 5-epi-alpha-selinene, (-)-delta-cadinene are found only in bryophytes, whereas (E)-2-epi-beta-caryophyllene is present only in lycophytes. Alpha-bisabolene, (+)-delta-selinene, alpha-longipinene are exclusive for acrogymnosperms. Bicyclogermacrene, 7-epi-sesquithujene, sesquithujene, beta-sesquiphellandrene are found only in monocots. Specific for dicots are gamma-muurolene, 7-epizingiberene, (Z)-gamma-bisabolene, germacrene-C, zingiberene, cis-muuroladiene, alpha-santalene, (+)-gamma-cadinene, and delta-guaiene.

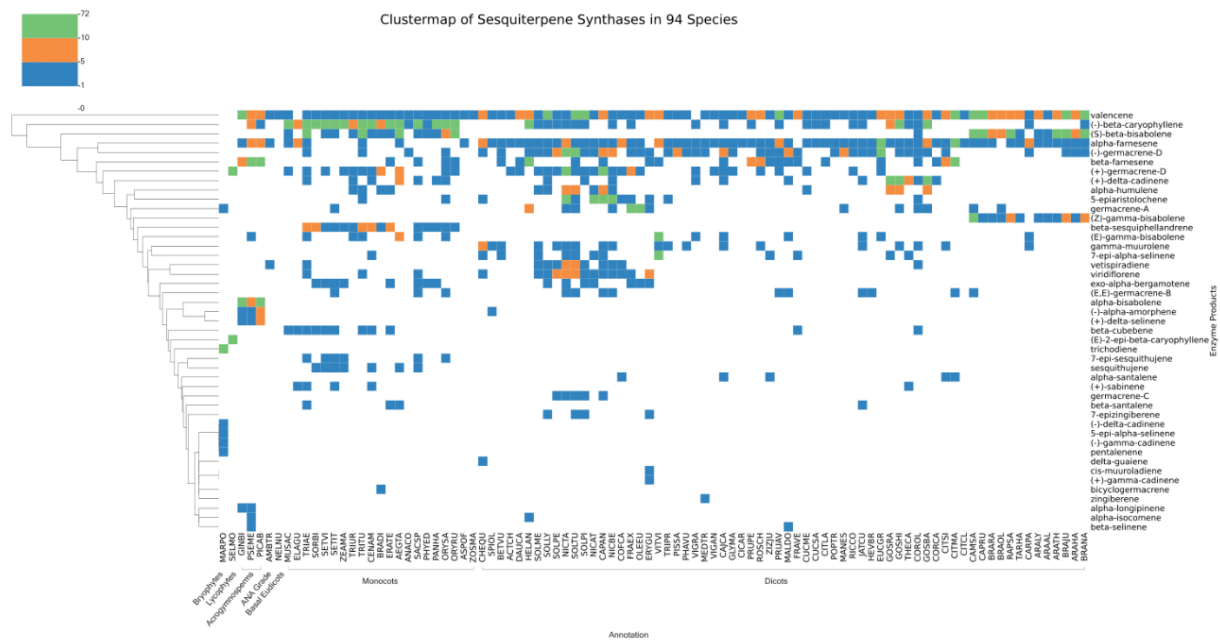

**Figure 4. Cluster map of sesquiterpene synthases in 94 species.** Sesquiterpene types are indicated in rows and species in columns, while the cell colors represent the number of given synthases in a species.

The results from E2P2 can then be further fed into Pathway Tools Software to predict the possible pathways in the plant [119]. Predicted pathways can subsequently be fed into the Semi-Automated Validation and Integration pipeline (SAVI), creating a final pathway in a species-specific database [120]. Taken together, the high abundance and diversity of the detected synthases indicate that many STLs remain to be discovered.

*Expression-based Pathway Predictions*
Secondary metabolites, including sesquiterpene lactones, are often expressed in specific organs and tissues [121,122]. Hence, the enzymes involved in the biosynthesis pathway and the mRNA encoding for these enzymes are also highly likely to be only present in the cells where the compounds of interest are high in quantity. Thus, the correlation of the gene expression pattern and the STL level can be used to predict which transcript is involved in the biosynthesis pathway. However, there are cases where the site of the synthesis and accumulation of the metabolites differ, such as nicotine, which is synthesized in roots but then transported to leaves [123,124]. Nevertheless, this analysis has successfully predicted the biosynthesis pathways of modified fatty acids in *Solanum lycopersicum* [125] and colchicine in *Gloriosa superba* [126]. Hence, adding to the criteria of strong candidates for STL genes is that the transcript has to be expressed highly only in the organs where the STL is accumulated.

The discovery of genes that are co-regulated in their transcription has allowed the completion of many metabolic pathways, including the production pathway of protolimonoids in *Azadirachta indica* [127], vinblastine in Madagascar periwinkle [128], etoposide glycone in *Podophyllum hexandrum* [129], and seco-iridoid in *Catharanthus roseus* [130]. Different methods can be utilized to identify these co-expressed genes. The most widely used approach is based on the "one versus all" analysis, where the expression profile of one gene is compared to those of the other genes, in which, afterward the expression profile similarity is sorted using metrics like the Pearson Correlation Coefficient (PCC; [131]). As a result, unknown components of biological processes could be discovered from the top 50 most relevant genes [132,133]. Another method is based on hierarchical clustering, a "many versus many" analysis. Genes are grouped into clusters based on their expression profile similarity, and visual analysis is used to identify relevant genes for a pathway. This method is useful when dealing with extensive lists of candidate genes. Additionally, co-expression networks, representing genes as nodes and connecting genes with similar expression profiles as edges, are employed for both "many versus many" and "all versus all" analyses. Despite differences from lists and hierarchical clustering, co-expression networks offer information about functionally related genes when used with a single query gene.

**Conclusion**

In summary, sesquiterpene lactones (STLs) are a diverse and biologically significant group of secondary metabolites, particularly prominent in the Asteraceae family. Their varied structures, derived from a 15-carbon backbone, contribute to various biological activities. STLs play crucial roles in plant defense against herbivory, acting as potent anti-herbivory agents and affecting insect development and mortality. They also function as phytotoxins, inhibiting the growth of competing plants, although their effectiveness in natural conditions remains debated. Furthermore, STLs serve as environmental signaling molecules, influencing the germination and growth of parasitic plant species and interacting with plant hormones, affecting processes like phototropism.

Beyond their ecological functions, STLs have significant medical applications. Artemisinin, a notable STL, is a crucial antimalarial drug, while others show promising anti-cancer and anti-inflammatory properties. The diverse biological activities of STLs are primarily attributed to their lactone structures, particularly the α-methylene-γ-lactone group, which enables the alkylation of thiol groups in proteins. However, this alkylation mechanism also accounts for some toxic properties.

The biosynthesis of STLs involves the mevalonic acid pathway and various modifications like oxidation, halogenation, acetylation, glycosylation, and glucosylation, leading to a wide array of structures. Understanding and predicting STL biosynthesis pathways are crucial for exploiting their therapeutic potential, necessitating advanced techniques like sequence-based and expression-based pathway predictions and co-expression analysis. This deep understanding of STLs' ecological functions, biosynthetic pathways, and medical relevance underscores their significance in plant biology and human health.